\documentclass[10pt]{article}

\usepackage{graphicx} 
\usepackage{color}
\usepackage{amssymb}

\begin{document}

\title{Hierarchical geometries and adhesion: Bio-inspired designs
for stiff interfaces}

\author{D. Rayneau-Kirkhope$^{1,2}$, Y. Mao$^3$, C. Rauch$^4$}

\maketitle
\begin{center}
$^1$ Department of Applied Physics, Aalto University, 02150 Espoo, FINLAND\\
  $^2$ Aalto Science Institute, School of Science, Aalto University, 02150 Espoo, FINLAND\\
$^3$ School of Physics and Astronomy, University of Nottingham, NG7 2RD, UK\\  
$^4$ School of Veterinary Medicine and Science, University of Nottingham, Sutton Bonington, LE12 5RD, UK
\end{center}

\begin{abstract}
Throughout biology, hierarchy is a recurrent theme in the geometry of structures where strength is achieved with minimal use of material. 
Acting over vast timescales, evolution has brought about beautiful solutions to problems of optimisation that are only now being understood and incorporated into engineering design. 
One particular example of this hierarchy is found in the junction between stiff keratinised material and the soft biological matter within the hooves of ungulates.
Using this biological interface as a design motif, we investigate the role of hierarchy in the creation of a stiff, robust interface between two materials. 
We show that through hierarchical design one can manipulate the scaling laws relating constituent material stiffness and overall interface stiffness under both shear and tension loading. 
Furthermore, we uncover a cascade of scaling laws for the higher order structure and link their origin with competing deformation modes within the structure. 
We demonstrate that when joining two materials of different stiffness, under shear or tension, hierarchical geometries are linked with beneficial mechanical properties. 
\end{abstract}

\section{Introduction}
Naturally occurring hierarchical interfaces for modulating adhesive interaction between two surfaces have been well documented; examples of such designs are to be found on the feet of geckos, spiders and insects \cite{Federle_2006, Gao_2006,Labonte_2016}. Although other contributions have been proposed \cite{Huber_2005}, it is understood that the primary interaction responsible for allowing geckos to walk up walls is the van der Waals interaction \cite{Autumn_2002}. The hierarchical geometry of the gecko's foot is key in making this adhesion possible utilising this very weak interaction \cite{Gao_2006b}. This structure has inspired a research area with the goal of creating dry adhesive mechanisms, with a high degree of success \cite{Geim_2003, Brodoceanu_2016, Cutkosky_2015, Autumn_2008}. Numerous other applications of fractal-like geometries are found in nature optimised for various functionalities including spider capture silk for strength and elasticity \cite{Zhou_2005}, hard biological composites for stiffness and fracture toughness \cite{Ritchie_2011, Li_2012} and trabecular bone for stiffness and minimal weight \cite{Huiskes_2000}. Excitingly, novel manufacturing methods are allowing the principles of hierarchical design to be utilised in engineering designs \cite{Rayneau-Kirkhope_2012, Rayneau-Kirkhope_2013, Schaelder_2011, Lin_2014}. 

The importance of geometry when creating an interface is clear: it is often the case that interfaces with non-trivial geometry exhibit remarkable mechanical properties \cite{Boyce_2013, Li_2014, Li_2012_a}. 
Suture joints are a prime example of such geometric specialization for mechanical purpose, typically observed joining two regions of a given material via an interfacial region comprised of a second material with a lower stiffness \cite{Boyce_2013, Shahar_2009}. 
Examples of such joints are to be found in bone \cite{Gao_2006} (including the cranium \cite{Hubbard_1971}), turtle shell \cite{Shahar_2009} and ammonites \cite{Kondo_2016, Boyce_2012}. In all of these cases, mechanical function is hypothesised to be a driving factor in the design of the interface (including sitffness, strength, fracture/penetration resistance); other functions for which such designs are specialised include growth, respiration and buoyancy control \cite{Boyce_2012, Kondo_2016, Hubbard_1971}.  

\begin{figure}
\begin{center}
\includegraphics[scale=0.7]{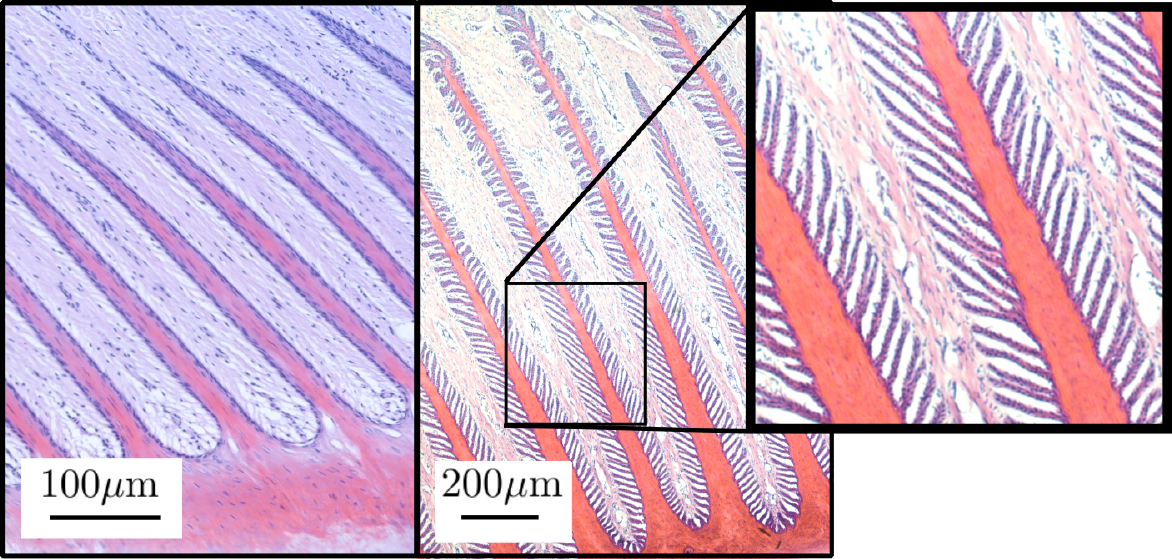}
\end{center}
\caption{\footnotesize [Colour online] A cross sectional image taken perpendicular to the surface of the hoof wall. The image shows the interface between soft biological material and stiff keretinized materials found in the bovine (left) and equine (right) hoof. In both cases the hoof wall is situated below the region imaged, while the pedal bone is above. This interface mediates large concussive loads between the two regions during the animal locomotion. The secondary lamellae are on both the dermal and epidermal lamellae are clearly visible in the inset of the right hand image. \label{bio_interface}}
\end{figure}

Here we focus on a novel application of hierarchical design - creating a permanent adhesive connection between two materials of vastly differing stiffness. 
The geometry of both sides of the interface is designed to permit maximal interface stiffness. 
The inspiration for our geometry stems from the interface between epidermal and dermal lamellae within the bovine and equine hoof capsule, shown in figure \ref{bio_interface}. 
It is of particular interest that different hierarchical orders of the same base motif can be seen in different species, thus giving the possibility of observing dependence of the optimal design on various input parameters.
In the equine hoof, this interface, long since known to exhibit a hierarchical geometry \cite{Stump_1967, Kainer_1989}, is responsible for the suspension of the weight of the horse from the hoof wall \cite{Pollitt_2004}. The geometry of this interface is assumed to be an anatomical specialisation key to force mediation \cite{Pollitt_1996a}. 
Clinical signs of laminitis (a disease causing lameness) occur when this hierarchical lamella architecture disintegrates \cite{Pollitt_1999}, thus heavy selection pressure should ensure that this interface create a robust connection between the pedal bone and the hoof wall \cite{Pollitt_2004, Pollitt_2010}. 
The role of the hierarchical geometry in the mediation of forces across this interface is currently not well understood. While the extensive surface area available for adhesion is used by many as justification for the ability of the horse foot to survive large load \cite{Pollitt_2004}, no studies have attempted to model this hypothesis.    

In this paper we present an investigation into the effect of hierarchical geometry on interface stiffness. 
We demonstrate the manipulation of scaling relationships relating overall interface stiffness to relative material stiffness through alteration of the hierarchical order. 
We show that under both shear and tension loading, beneficial properties are associated with increasing the hierarchical order of the structure. 
In higher order structures, we uncover a cascade of scaling relationships that are linked to alternative deformation modes within the interface structure. 
We show the transition from one scaling regime to another can be manipulated through alteration of the geometric parameters within the interface geometry.

\section{Geometry and hierarchy}

\begin{figure}
\begin{center}
\raisebox{0.5cm}{\resizebox{5cm}{!}{\begingroup%
  \makeatletter%
  \providecommand\color[2][]{%
    \errmessage{(Inkscape) Color is used for the text in Inkscape, but the package 'color.sty' is not loaded}%
    \renewcommand\color[2][]{}%
  }%
  \providecommand\transparent[1]{%
    \errmessage{(Inkscape) Transparency is used (non-zero) for the text in Inkscape, but the package 'transparent.sty' is not loaded}%
    \renewcommand\transparent[1]{}%
  }%
  \providecommand\rotatebox[2]{#2}%
  \ifx\svgwidth\undefined%
    \setlength{\unitlength}{134.47349969bp}%
    \ifx\svgscale\undefined%
      \relax%
    \else%
      \setlength{\unitlength}{\unitlength * \real{\svgscale}}%
    \fi%
  \else%
    \setlength{\unitlength}{\svgwidth}%
  \fi%
  \global\let\svgwidth\undefined%
  \global\let\svgscale\undefined%
  \makeatother%
  \begin{picture}(1,0.40225619)%
    \put(0,0){\includegraphics[width=\unitlength,page=1]{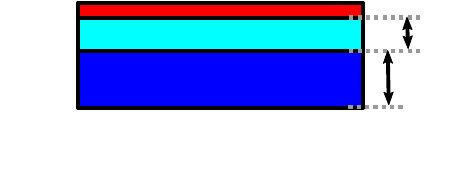}}%
    \put(0.9087267,0.32271505){\color[rgb]{0,0,0}\makebox(0,0)[lb]{\smash{$t_{i}$}}}%
    \put(-0.002045,0.17611066){\color[rgb]{0,0,0}\makebox(0,0)[lb]{\smash{$Y_{s}$}}}%
    \put(0.04162357,0.37442133){\color[rgb]{0,0,0}\makebox(0,0)[lb]{\smash{$Y_{i}$}}}%
    \put(0.85390642,0.22447383){\color[rgb]{0,0,0}\makebox(0,0)[lb]{\smash{$t_{s}$}}}%
    \put(0,0){\includegraphics[width=\unitlength,page=2]{G0_interface_2.pdf}}%
    \put(0.45006618,-0.00334609){\color[rgb]{0,0,0}\makebox(0,0)[lb]{\smash{$L_x$}}}%
    \put(0,0){\includegraphics[width=\unitlength,page=3]{G0_interface_2.pdf}}%
    \put(0.40589616,0.07237384){\color[rgb]{0,0,0}\makebox(0,0)[lb]{\smash{$\Delta$}}}%
  \end{picture}%
\endgroup%
}}\\
\raisebox{0.2cm}{\resizebox{6cm}{!}{\begingroup%
  \makeatletter%
  \providecommand\color[2][]{%
    \errmessage{(Inkscape) Color is used for the text in Inkscape, but the package 'color.sty' is not loaded}%
    \renewcommand\color[2][]{}%
  }%
  \providecommand\transparent[1]{%
    \errmessage{(Inkscape) Transparency is used (non-zero) for the text in Inkscape, but the package 'transparent.sty' is not loaded}%
    \renewcommand\transparent[1]{}%
  }%
  \providecommand\rotatebox[2]{#2}%
  \ifx\svgwidth\undefined%
    \setlength{\unitlength}{161.22959888bp}%
    \ifx\svgscale\undefined%
      \relax%
    \else%
      \setlength{\unitlength}{\unitlength * \real{\svgscale}}%
    \fi%
  \else%
    \setlength{\unitlength}{\svgwidth}%
  \fi%
  \global\let\svgwidth\undefined%
  \global\let\svgscale\undefined%
  \makeatother%
  \begin{picture}(1,0.78211347)%
    \put(0,0){\includegraphics[width=\unitlength,page=1]{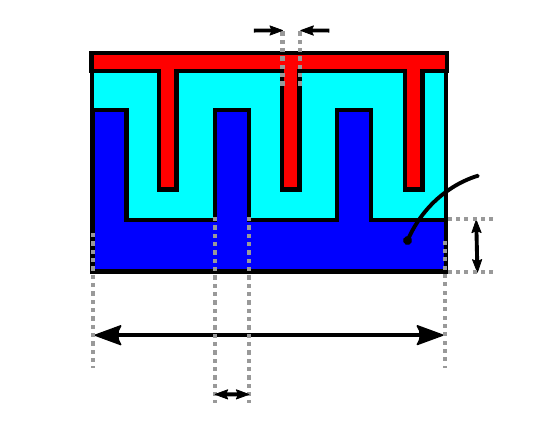}}%
    \put(0.5380638,0.75273116){\color[rgb]{0,0,0}\makebox(0,0)[lb]{\smash{$t_{F}$}}}%
    \put(0.85574692,0.45663205){\color[rgb]{0,0,0}\makebox(0,0)[lb]{\smash{$Y_{s}$}}}%
    \put(0.0307137,0.7240764){\color[rgb]{0,0,0}\makebox(0,0)[lb]{\smash{$Y_{i}$}}}%
    \put(0.87261739,0.33109715){\color[rgb]{0,0,0}\makebox(0,0)[lb]{\smash{$t_{s}$}}}%
    \put(0.39776716,0.01552276){\color[rgb]{0,0,0}\makebox(0,0)[lb]{\smash{$t_{1,1}$}}}%
    \put(0,0){\includegraphics[width=\unitlength,page=2]{G1_interface.pdf}}%
    \put(0.50692826,0.12617206){\color[rgb]{0,0,0}\makebox(0,0)[lb]{\smash{$L_{x}$}}}%
    \put(0,0){\includegraphics[width=\unitlength,page=3]{G1_interface.pdf}}%
    \put(-0.00215868,0.47437079){\color[rgb]{0,0,0}\makebox(0,0)[lb]{\smash{$h_{1,1}$}}}%
    \put(0,0){\includegraphics[width=\unitlength,page=4]{G1_interface.pdf}}%
    \put(0.87261739,0.60896176){\color[rgb]{0,0,0}\makebox(0,0)[lb]{\smash{$t_{i}$}}}%
    \put(0,0){\includegraphics[width=\unitlength,page=5]{G1_interface.pdf}}%
    \put(0.26875859,0.21548568){\color[rgb]{0,0,0}\makebox(0,0)[lb]{\smash{$\Delta$}}}%
    \put(0,0){\includegraphics[width=\unitlength,page=6]{G1_interface.pdf}}%
    \put(0.14967376,0.00311809){\color[rgb]{0,0,0}\makebox(0,0)[lb]{\smash{$x$}}}%
    \put(0.04795547,0.09863405){\color[rgb]{0,0,0}\makebox(0,0)[lb]{\smash{$y$}}}%
    \put(-0.18743526,0.17738679){\color[rgb]{0,0,0}\makebox(0,0)[lt]{\begin{minipage}{0.30701561\unitlength}\raggedright \end{minipage}}}%
  \end{picture}%
\endgroup%
}}\\
\resizebox{7cm}{!}{\begingroup%
  \makeatletter%
  \providecommand\color[2][]{%
    \errmessage{(Inkscape) Color is used for the text in Inkscape, but the package 'color.sty' is not loaded}%
    \renewcommand\color[2][]{}%
  }%
  \providecommand\transparent[1]{%
    \errmessage{(Inkscape) Transparency is used (non-zero) for the text in Inkscape, but the package 'transparent.sty' is not loaded}%
    \renewcommand\transparent[1]{}%
  }%
  \providecommand\rotatebox[2]{#2}%
  \ifx\svgwidth\undefined%
    \setlength{\unitlength}{217.06856315bp}%
    \ifx\svgscale\undefined%
      \relax%
    \else%
      \setlength{\unitlength}{\unitlength * \real{\svgscale}}%
    \fi%
  \else%
    \setlength{\unitlength}{\svgwidth}%
  \fi%
  \global\let\svgwidth\undefined%
  \global\let\svgscale\undefined%
  \makeatother%
  \begin{picture}(1,0.70193729)%
    \put(0,0){\includegraphics[width=\unitlength,page=1]{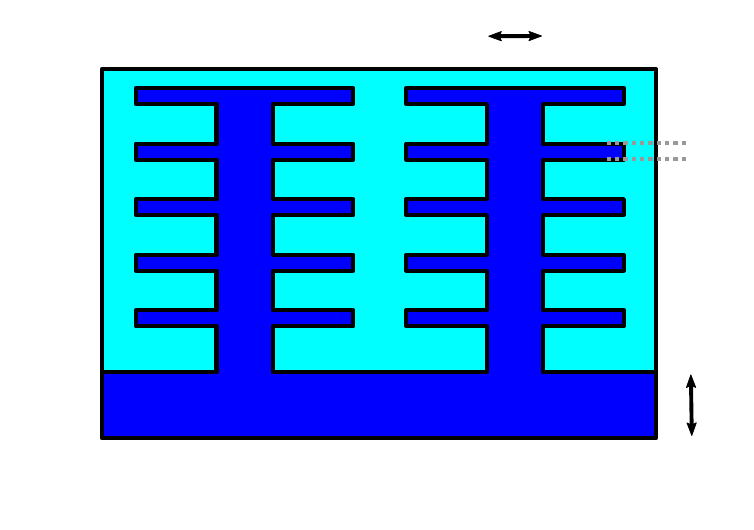}}%
    \put(0.33376958,0.52562249){\color[rgb]{1,1,1}\makebox(0,0)[lb]{\smash{$\theta$}}}%
    \put(0.0466094,0.13757275){\color[rgb]{0,0,0}\makebox(0,0)[lb]{\smash{$Y_{s}$}}}%
    \put(0.90534765,0.52575768){\color[rgb]{0,0,0}\makebox(0,0)[lb]{\smash{$t_{2,1}$}}}%
    \put(0,0){\includegraphics[width=\unitlength,page=2]{G2_interface.pdf}}%
    \put(0.3073799,0.06415237){\color[rgb]{0,0,0}\makebox(0,0)[lb]{\smash{$\Delta$}}}%
    \put(0,0){\includegraphics[width=\unitlength,page=3]{G2_interface.pdf}}%
    \put(0.66210653,0.68469364){\color[rgb]{0,0,0}\makebox(0,0)[lb]{\smash{$t_{2,2}$}}}%
    \put(0.91271859,0.24566185){\color[rgb]{0,0,0}\makebox(0,0)[lb]{\smash{$t_{i}$}}}%
    \put(0.91271859,0.41519353){\color[rgb]{0,0,0}\makebox(0,0)[lb]{\smash{$t_{F}$}}}%
    \put(0,0){\includegraphics[width=\unitlength,page=4]{G2_interface.pdf}}%
    \put(0.18843992,0.65688752){\color[rgb]{0,0,0}\makebox(0,0)[lb]{\smash{$Y_{i}$}}}%
    \put(0,0){\includegraphics[width=\unitlength,page=5]{G2_interface.pdf}}%
    \put(0.92400535,0.16227805){\color[rgb]{0,0,0}\makebox(0,0)[lb]{\smash{$t_{s}$}}}%
    \put(0,0){\includegraphics[width=\unitlength,page=6]{G2_interface.pdf}}%
    \put(-0.00126686,0.4020231){\color[rgb]{0,0,0}\makebox(0,0)[lb]{\smash{$l_{2,2}$}}}%
    \put(0,0){\includegraphics[width=\unitlength,page=7]{G2_interface.pdf}}%
    \put(0.38076429,0.68270064){\color[rgb]{0,0,0}\makebox(0,0)[lb]{\smash{$l_{2,1}$}}}%
    \put(0.03221303,0.44975765){\color[rgb]{0,0,0}\makebox(0,0)[lb]{\smash{}}}%
    \put(0,0){\includegraphics[width=\unitlength,page=8]{G2_interface.pdf}}%
    \put(0.50409194,0.05309596){\color[rgb]{0,0,0}\makebox(0,0)[lb]{\smash{$L_{x}$}}}%
  \end{picture}%
\endgroup%
}
\caption{\footnotesize [Colour online] Top: the generation-0 geometry: a planar connection of two materials of vastly differing stiffness joined by an adhesive connection of thickness $t_i$. Middle: the generation-1 interface and its parameterisation. Bottom: the generation-2 geometry: A pair of primary lamellae protrude from the deformable domain, from these primary lamellae, a secondary set of lamellae emanate at a given angle, $\theta$ ($\theta = \pi/2$ shown). The fixed domain interdigitates these deformable lamellae. In all three diagrams, the red region represents an infinitely stiff material, which is connected to the deformable material (shown in blue) via a (turquoise) inter-surface interaction. An imposed displacement of $\Delta$ is applied to the lower surface of a given design and the stiffness of the interface is evaluated.}\label{geometry}
\end{center}
\end{figure}

The geometry of the interface investigated here is simple to construct in an iterative manner. First, we define the generation-0 surface as a simple planar interface, see figure \ref{geometry}. 
The generation-1 structure can be thought of as a generation-0 structure with a series of lamellae protruding from both sides of the interface creating an interlocking pattern, the resulting geometry is shown in figure \ref{geometry} where the parameterisation of the geometry is introduced. 
The generation-$n$ structure can be constructed by placing a set of interlocking lamella protruding from both surfaces of the generation-($n-1$) geometry normal to the tangent of the interface at that point, at set interval along the arc-length of the interface. At each iteration it is assumed that all lamellae introduced are of the same geometry.
A pair of primary lamella of a generation-2 surface is shown in figure \ref{geometry}. 

Here we focus on joining two materials with vastly differing stiffness, thus we model one side of the interface as infinitely rigid, while the second is permitted to deform with a linear elastic response. Between the two surfaces, a third material is added which is assumed to be Hookean and isotropic. 
The length and width of the lamellae is defined as $l_{G,i}$ and $t_{G,i}$ respectively where $G$ is the generation of the structure and $i$ is iteration at which the lamellae were introduced of the lamellae ($i=1$ corresponds to the first lamellae protrude from the deformable domain base, $i=2$ corresponds to the lamellae that protrude from the $i=1$ lamellae). 
The aspect ratio of each lamella is then defined as 
\begin{equation}
a_{G,i} \equiv \frac{l_{G,i}}{t_{G,i}}. \label{aspect}
\end{equation}
In order to test the stiffness of the interface an imposed displacement of magnitude $\Delta$ is applied to the lower surface of the deformable material and the reaction force is measured along that boundary. 

\section{Results}\label{Res}

Here we present the results of finite element simulations calculating the stiffness of geometries with varying degrees of hierarchy.  
The finite element simulations are undertaken using the two-dimensional structural mechanics module of COMSOL 5.1 Multiphysics \cite{COMSOL} using a plane strain assumption. Mesh refinement studies were undertaken to ascertain accuracy of the results, required mesh density was highly dependent on the relative thickness of interface to lamellae and other parameters. 

\vspace{5mm}
\subsection{Generation-0}
\begin{figure}
\begin{center}
\resizebox{6cm}{!}{\begingroup%
  \makeatletter%
  \providecommand\color[2][]{%
    \errmessage{(Inkscape) Color is used for the text in Inkscape, but the package 'color.sty' is not loaded}%
    \renewcommand\color[2][]{}%
  }%
  \providecommand\transparent[1]{%
    \errmessage{(Inkscape) Transparency is used (non-zero) for the text in Inkscape, but the package 'transparent.sty' is not loaded}%
    \renewcommand\transparent[1]{}%
  }%
  \providecommand\rotatebox[2]{#2}%
  \ifx\svgwidth\undefined%
    \setlength{\unitlength}{229.1843204bp}%
    \ifx\svgscale\undefined%
      \relax%
    \else%
      \setlength{\unitlength}{\unitlength * \real{\svgscale}}%
    \fi%
  \else%
    \setlength{\unitlength}{\svgwidth}%
  \fi%
  \global\let\svgwidth\undefined%
  \global\let\svgscale\undefined%
  \makeatother%
  \begin{picture}(1,0.64745587)%
    \put(0,0){\includegraphics[width=\unitlength,page=1]{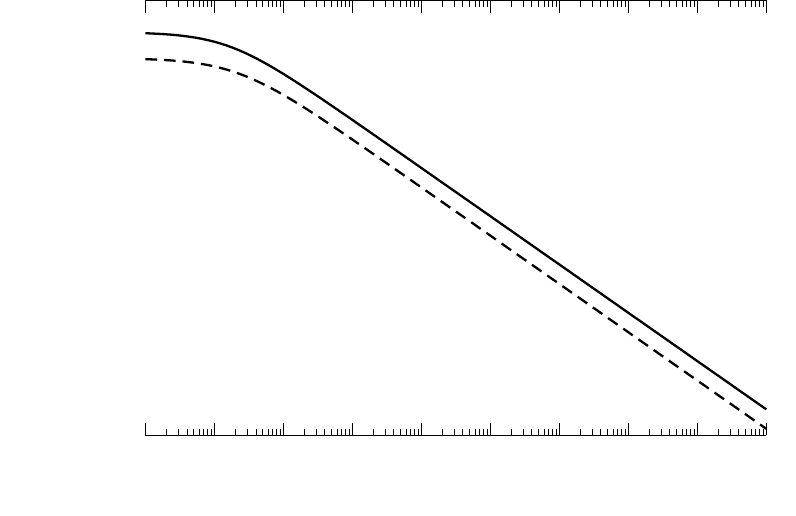}}%
    \put(0.17849344,0.05031668){\color[rgb]{0,0,0}\makebox(0,0)[lb]{\smash{1}}}%
    \put(0.25877872,0.05031668){\color[rgb]{0,0,0}\makebox(0,0)[lb]{\smash{10}}}%
    \put(0.3400783,0.05031668){\color[rgb]{0,0,0}\makebox(0,0)[lb]{\smash{100}}}%
    \put(0.42129985,0.05031668){\color[rgb]{0,0,0}\makebox(0,0)[lb]{\smash{$10^3$}}}%
    \put(0.50244338,0.05031668){\color[rgb]{0,0,0}\makebox(0,0)[lb]{\smash{$10^4$}}}%
    \put(0.58959465,0.05000459){\color[rgb]{0,0,0}\makebox(0,0)[lb]{\smash{$10^5$}}}%
    \put(0.67604372,0.05000459){\color[rgb]{0,0,0}\makebox(0,0)[lb]{\smash{$10^6$}}}%
    \put(0.7628829,0.05031668){\color[rgb]{0,0,0}\makebox(0,0)[lb]{\smash{$10^7$}}}%
    \put(0.84956604,0.05031668){\color[rgb]{0,0,0}\makebox(0,0)[lb]{\smash{$10^8$}}}%
    \put(0.93609313,0.05031668){\color[rgb]{0,0,0}\makebox(0,0)[lb]{\smash{$10^9$}}}%
    \put(0,0){\includegraphics[width=\unitlength,page=2]{G0_both.pdf}}%
    \put(0.06083782,0.08655295){\color[rgb]{0,0,0}\makebox(0,0)[lb]{\smash{$10^{-7}$}}}%
    \put(0.06083782,0.1470985){\color[rgb]{0,0,0}\makebox(0,0)[lb]{\smash{$10^{-6}$}}}%
    \put(0.06083782,0.20772208){\color[rgb]{0,0,0}\makebox(0,0)[lb]{\smash{$10^{-5}$}}}%
    \put(0.06083782,0.26857973){\color[rgb]{0,0,0}\makebox(0,0)[lb]{\smash{$10^{-4}$}}}%
    \put(0.06083782,0.32928133){\color[rgb]{0,0,0}\makebox(0,0)[lb]{\smash{$10^{-3}$}}}%
    \put(0.06935723,0.38998294){\color[rgb]{0,0,0}\makebox(0,0)[lb]{\smash{0.01}}}%
    \put(0.09400889,0.45068454){\color[rgb]{0,0,0}\makebox(0,0)[lb]{\smash{0.1}}}%
    \put(0.12451224,0.51115207){\color[rgb]{0,0,0}\makebox(0,0)[lb]{\smash{1}}}%
    \put(0.1116385,0.57200972){\color[rgb]{0,0,0}\makebox(0,0)[lb]{\smash{10}}}%
    \put(0.08698685,0.63271132){\color[rgb]{0,0,0}\makebox(0,0)[lb]{\smash{100}}}%
    \put(0,0){\includegraphics[width=\unitlength,page=3]{G0_both.pdf}}%
    \put(0.74315367,0.57809549){\color[rgb]{0,0,0}\makebox(0,0)[lb]{\smash{Shear}}}%
    \put(0,0){\includegraphics[width=\unitlength,page=4]{G0_both.pdf}}%
    \put(0.74315367,0.53405696){\color[rgb]{0,0,0}\makebox(0,0)[lb]{\smash{Tension}}}%
    \put(0,0){\includegraphics[width=\unitlength,page=5]{G0_both.pdf}}%
    \put(0.52309059,0.00199414){\color[rgb]{0,0,0}\makebox(0,0)[lb]{\smash{$\eta$}}}%
    \put(-0.00138054,0.3593151){\color[rgb]{0,0,0}\makebox(0,0)[lb]{\smash{$f_R$}}}%
    \put(0,0){\includegraphics[width=\unitlength,page=6]{G0_both.pdf}}%
    \put(0.70085153,0.37703611){\color[rgb]{0,0,0}\makebox(0,0)[lb]{\smash{1}}}%
  \end{picture}%
\endgroup%
}
\caption{\footnotesize The stiffness of a generation-0 interface. The reaction force shown is for a structure with parameters $L_x = 3\times 10^{-4}$m $t_i=1\times 10^{-6}$m, $Y_i = 100$MPa, $\Delta = \frac{t_i}{10}$. }\label{G0_res}
\end{center}
\end{figure}

The generation-0 structure is composed of a planar interface between a deformable and an infinitely rigid material connected by an intermediate elastic medium, see figure \ref{geometry}. 
When imposing a displacement on the base of the deformable surface relative to the rigid material, the nature of the deformation across the structure will be dependent on the stiffness of the interface, $Y_i$, and that of the deformable surface, $Y_s$, the ratio of the stiffness will be denoted $\eta$:
\begin{equation}
\eta = \frac{Y_s}{Y_i}.
\end{equation} 
We also introduce the non-dimensional load parameter, 
\begin{equation}
f_R(G) \equiv \frac{F_R(G)}{Y_s \Delta },
\end{equation}
where $\Delta$ is the imposed displacement on the structure, $G$ denotes the generation of the structure investigated and $F_R$ is the reaction load parallel to the imposed displacement per unit length in the remaining spatial dimension. 
Assuming periodic boundary conditions in the $x$ direction, it is possible to show that the minimum energy configuration under loading preserves parallel lines perpendicular to the normal of the interface. 
As shown in \ref{G0_res}, the reaction forces for the flat interface under tension and shear both follow the same scaling behaviour with $\eta$: For small $\eta$ the reaction force measured on the interface is independent of variations in $\eta$, while for larger $\eta$ the scaling observed is
\begin{equation}
f_R \sim \eta^{-1}.\label{G0_scaling}
\end{equation}

\vspace{5mm}

\subsection{Generation-1}
For the generation-1 and higher order designs, we introduce $\gamma$ as the ratio of the total reaction force of the structure of interest to that of the generation-0 structure of the same width ($L_x$ in figure \ref{geometry}) undergoing the same deformation,
\begin{equation}
\gamma \equiv \frac{f_R(G)}{f_R(0)}.
\end{equation}
As shown in figure \ref{geometry}, the generation one interface is made up of a flat interface between a deformable (with Young's Modulus $Y_s$) and infinitely stiff material with a series of interdigitated lamellae protruding perpendicular to the interface from either side. 
The two materials are joined by a third material with Young's Modulus $Y_i$. 
The parameters used in obtaining the below results are $ t_{1,1} = 5\times 10 ^{-4}$m, $ t_i = 1\times 10^{-6}$m, $ t_s = 5\times 10^{-5}$m, $t_F = t_{1,1},$ $Y_i = 1\times 10^8$Pa, $\Delta = t_i/10$ and $\nu = 0.45$.

\begin{figure}
\begin{center}
\resizebox{6cm}{!}{\begingroup%
  \makeatletter%
  \providecommand\color[2][]{%
    \errmessage{(Inkscape) Color is used for the text in Inkscape, but the package 'color.sty' is not loaded}%
    \renewcommand\color[2][]{}%
  }%
  \providecommand\transparent[1]{%
    \errmessage{(Inkscape) Transparency is used (non-zero) for the text in Inkscape, but the package 'transparent.sty' is not loaded}%
    \renewcommand\transparent[1]{}%
  }%
  \providecommand\rotatebox[2]{#2}%
  \ifx\svgwidth\undefined%
    \setlength{\unitlength}{221.67525164bp}%
    \ifx\svgscale\undefined%
      \relax%
    \else%
      \setlength{\unitlength}{\unitlength * \real{\svgscale}}%
    \fi%
  \else%
    \setlength{\unitlength}{\svgwidth}%
  \fi%
  \global\let\svgwidth\undefined%
  \global\let\svgscale\undefined%
  \makeatother%
  \begin{picture}(1,0.7204624)%
    \put(0,0){\includegraphics[width=\unitlength,page=1]{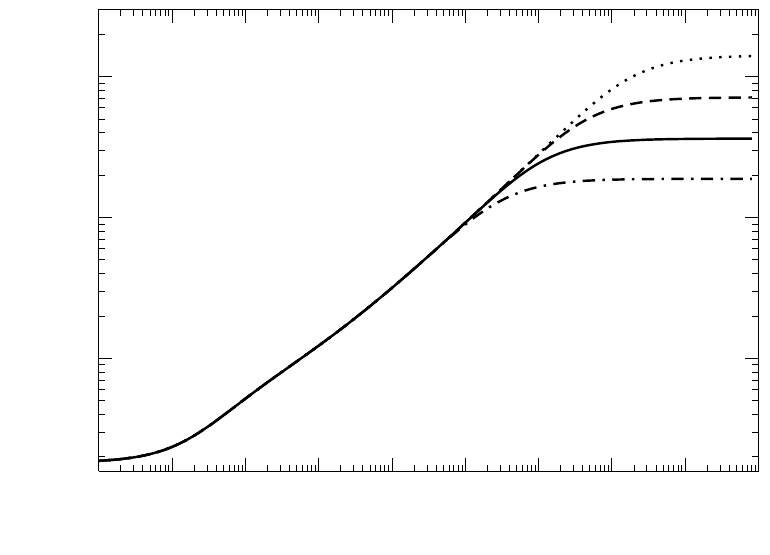}}%
    \put(0.0959956,0.24671257){\color[rgb]{0,0,0}\makebox(0,0)[lb]{\smash{1}}}%
    \put(0.07463561,0.42996156){\color[rgb]{0,0,0}\makebox(0,0)[lb]{\smash{10}}}%
    \put(0.05567551,0.61303912){\color[rgb]{0,0,0}\makebox(0,0)[lb]{\smash{100}}}%
    \put(0,0){\includegraphics[width=\unitlength,page=2]{G1_tension_amplification.pdf}}%
    \put(0.25319847,0.63940213){\color[rgb]{0,0,0}\makebox(0,0)[lb]{\smash{$a_{1,1} = 32$}}}%
    \put(0,0){\includegraphics[width=\unitlength,page=3]{G1_tension_amplification.pdf}}%
    \put(0.25319847,0.59988376){\color[rgb]{0,0,0}\makebox(0,0)[lb]{\smash{$a_{1,1} = 64$}}}%
    \put(0,0){\includegraphics[width=\unitlength,page=4]{G1_tension_amplification.pdf}}%
    \put(0.25319847,0.55280403){\color[rgb]{0,0,0}\makebox(0,0)[lb]{\smash{$a_{1,1} = 128$}}}%
    \put(0,0){\includegraphics[width=\unitlength,page=5]{G1_tension_amplification.pdf}}%
    \put(0.25319847,0.505982){\color[rgb]{0,0,0}\makebox(0,0)[lb]{\smash{$a_{1,1} = 256$}}}%
    \put(0,0){\includegraphics[width=\unitlength,page=6]{G1_tension_amplification.pdf}}%
    \put(0.56496863,0.00206171){\color[rgb]{0,0,0}\makebox(0,0)[lb]{\smash{$\eta$}}}%
    \put(-0.00111014,0.41295856){\color[rgb]{0,0,0}\makebox(0,0)[lb]{\smash{$\gamma$}}}%
    \put(0.1233822,0.04496226){\color[rgb]{0,0,0}\makebox(0,0)[lb]{\smash{1}}}%
    \put(0.20868098,0.04496226){\color[rgb]{0,0,0}\makebox(0,0)[lb]{\smash{10}}}%
    \put(0.29509655,0.04496226){\color[rgb]{0,0,0}\makebox(0,0)[lb]{\smash{100}}}%
    \put(0.38142614,0.04496226){\color[rgb]{0,0,0}\makebox(0,0)[lb]{\smash{$10^3$}}}%
    \put(0.46775581,0.04496226){\color[rgb]{0,0,0}\makebox(0,0)[lb]{\smash{$10^4$}}}%
    \put(0.5606138,0.04461866){\color[rgb]{0,0,0}\makebox(0,0)[lb]{\smash{$10^5$}}}%
    \put(0.6526128,0.04461866){\color[rgb]{0,0,0}\makebox(0,0)[lb]{\smash{$10^6$}}}%
    \put(0.74512728,0.04496226){\color[rgb]{0,0,0}\makebox(0,0)[lb]{\smash{$10^7$}}}%
    \put(0.83746988,0.04496226){\color[rgb]{0,0,0}\makebox(0,0)[lb]{\smash{$10^8$}}}%
    \put(0.92964076,0.04496226){\color[rgb]{0,0,0}\makebox(0,0)[lb]{\smash{$10^9$}}}%
    \put(-0.00111014,0.7016691){\color[rgb]{0,0,0}\makebox(0,0)[lb]{\smash{(a)}}}%
    \put(0,0){\includegraphics[width=\unitlength,page=7]{G1_tension_amplification.pdf}}%
    \put(0.46212404,0.42857297){\color[rgb]{0,0,0}\makebox(0,0)[lb]{\smash{0.5}}}%
  \end{picture}%
\endgroup%
}
\resizebox{6cm}{!}{\begingroup%
  \makeatletter%
  \providecommand\color[2][]{%
    \errmessage{(Inkscape) Color is used for the text in Inkscape, but the package 'color.sty' is not loaded}%
    \renewcommand\color[2][]{}%
  }%
  \providecommand\transparent[1]{%
    \errmessage{(Inkscape) Transparency is used (non-zero) for the text in Inkscape, but the package 'transparent.sty' is not loaded}%
    \renewcommand\transparent[1]{}%
  }%
  \providecommand\rotatebox[2]{#2}%
  \ifx\svgwidth\undefined%
    \setlength{\unitlength}{222.17649205bp}%
    \ifx\svgscale\undefined%
      \relax%
    \else%
      \setlength{\unitlength}{\unitlength * \real{\svgscale}}%
    \fi%
  \else%
    \setlength{\unitlength}{\svgwidth}%
  \fi%
  \global\let\svgwidth\undefined%
  \global\let\svgscale\undefined%
  \makeatother%
  \begin{picture}(1,0.72012301)%
    \put(0,0){\includegraphics[width=\unitlength,page=1]{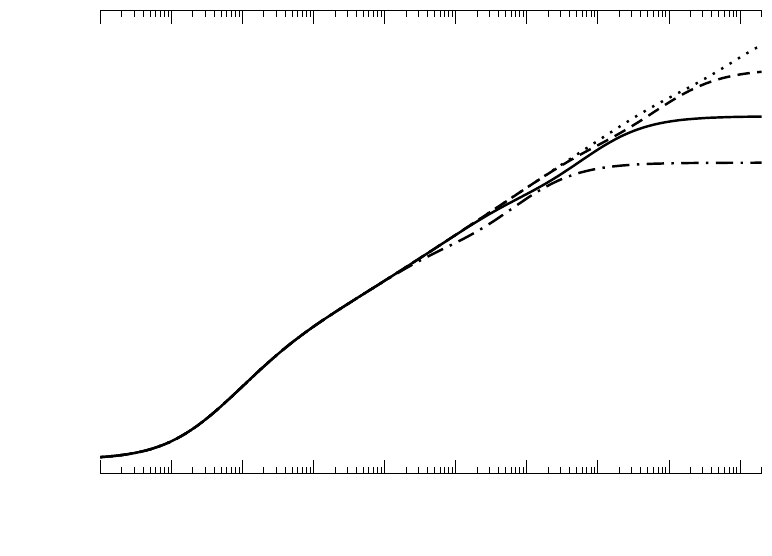}}%
    \put(0.12535943,0.04614679){\color[rgb]{0,0,0}\makebox(0,0)[lb]{\smash{1}}}%
    \put(0.21046586,0.04614679){\color[rgb]{0,0,0}\makebox(0,0)[lb]{\smash{10}}}%
    \put(0.29668647,0.04614679){\color[rgb]{0,0,0}\makebox(0,0)[lb]{\smash{100}}}%
    \put(0.38282138,0.04614679){\color[rgb]{0,0,0}\makebox(0,0)[lb]{\smash{$10^3$}}}%
    \put(0.46895628,0.04614679){\color[rgb]{0,0,0}\makebox(0,0)[lb]{\smash{$10^4$}}}%
    \put(0.56160487,0.04580396){\color[rgb]{0,0,0}\makebox(0,0)[lb]{\smash{$10^5$}}}%
    \put(0.6533964,0.04580396){\color[rgb]{0,0,0}\makebox(0,0)[lb]{\smash{$10^6$}}}%
    \put(0.74570216,0.04614679){\color[rgb]{0,0,0}\makebox(0,0)[lb]{\smash{$10^7$}}}%
    \put(0.83783652,0.04614679){\color[rgb]{0,0,0}\makebox(0,0)[lb]{\smash{$10^8$}}}%
    \put(0.92979946,0.04614679){\color[rgb]{0,0,0}\makebox(0,0)[lb]{\smash{$10^9$}}}%
    \put(0,0){\includegraphics[width=\unitlength,page=2]{G1_shear_amplification.pdf}}%
    \put(0.08326899,0.18217352){\color[rgb]{0,0,0}\makebox(0,0)[lb]{\smash{1}}}%
    \put(0.06912744,0.39086854){\color[rgb]{0,0,0}\makebox(0,0)[lb]{\smash{10}}}%
    \put(0.05738567,0.59939215){\color[rgb]{0,0,0}\makebox(0,0)[lb]{\smash{100}}}%
    \put(0,0){\includegraphics[width=\unitlength,page=3]{G1_shear_amplification.pdf}}%
    \put(0.26228816,0.63924561){\color[rgb]{0,0,0}\makebox(0,0)[lb]{\smash{$a_{1,1} = 4$}}}%
    \put(0,0){\includegraphics[width=\unitlength,page=4]{G1_shear_amplification.pdf}}%
    \put(0.26228816,0.5998164){\color[rgb]{0,0,0}\makebox(0,0)[lb]{\smash{$a_{1,1} = 8$}}}%
    \put(0,0){\includegraphics[width=\unitlength,page=5]{G1_shear_amplification.pdf}}%
    \put(0.26228816,0.55284287){\color[rgb]{0,0,0}\makebox(0,0)[lb]{\smash{$a_{1,1} = 16$}}}%
    \put(0,0){\includegraphics[width=\unitlength,page=6]{G1_shear_amplification.pdf}}%
    \put(0.26228816,0.50612647){\color[rgb]{0,0,0}\makebox(0,0)[lb]{\smash{$a_{1,1} = 32$}}}%
    \put(0,0){\includegraphics[width=\unitlength,page=7]{G1_shear_amplification.pdf}}%
    \put(0.55283263,0.00205704){\color[rgb]{0,0,0}\makebox(0,0)[lb]{\smash{$\eta$}}}%
    \put(-0.00142413,0.41331289){\color[rgb]{0,0,0}\makebox(0,0)[lb]{\smash{$\gamma$}}}%
    \put(0.00114793,0.7013721){\color[rgb]{0,0,0}\makebox(0,0)[lb]{\smash{(b)}}}%
    \put(0,0){\includegraphics[width=\unitlength,page=8]{G1_shear_amplification.pdf}}%
    \put(0.44636216,0.42889208){\color[rgb]{0,0,0}\makebox(0,0)[lb]{\smash{0.24}}}%
  \end{picture}%
\endgroup%
}
\caption{\footnotesize The stiffness increase of a generation-1 interface relative to a flat geometry (generation-0) as a function of $\eta$ for various aspect ratios of lamellae in the generation-1 design. The interface loaded under (a) tension and (b) shear. The figures show the results for various aspect ratios of structure, $a_{1,1}$, as defined in Eq.~(\ref{aspect}).}\label{G1_res}
\end{center}
\end{figure}

{\it Tension:}
We first consider tension imposed on the surface of the deformable surface: 
a displacement in the $-y$ direction is imposed along the base of the deformable material, as indicated by the yellow dashed line in figure \ref{geometry}, and the reaction force on the structure is measured. 
Figure \ref{G1_res}(a) shows the ratio of $f_R$ for the generation-1 and generation-0 structures of equal width (defined as $\gamma$) for varying $\eta$. 
It is observed that $\gamma$ varies non-trivially with $\eta$:
For $\eta \lesssim 10^2$ the generation-0 structure is stiffer than the generation-1 structure ($\gamma < 1$). 
For larger $\eta$, we see the generation-1 structure increases in stiffness relative to the generation-0 geometry, it is observed that
\begin{equation}
\gamma \sim \eta^{0.5},
\end{equation}
where the error in the power is $\pm 0.01$. 
For larger $\eta$ and finite aspect ratio, $\gamma$ is seen to be independent of $\eta$.

{\it Shear:}
We also investigate the effect of an imposed displacement in the $x$-direction: shear on the interface. As shown in figure \ref{G1_res}(b), the scaling of $\gamma$ with $\eta$ is described by a new power law: for sufficiently long lamellae and large $\eta$ it is observed that,
\begin{equation}
\gamma \sim \eta^{0.24},
\end{equation}
where the error in the power is $\pm 0.02$. 
As with the case of tension being imposed on the interface, for finite aspect ratio, above a certain value of $\eta$, $\gamma$ is observed to be independent of $\eta$. 

In both loading conditions, the transition in scaling of $\gamma$ with $\eta$ is due to the nature of the deformation within the structure: for values of $\eta$ where $\gamma$ is independent of $\eta$, the displacement at the tip of the lamellae is approximately equal to the displacement at the base of the lamella, thus the whole of the lamella is displaced as one rigid body. Consequently, the values of $f_R(0)$ and $f_R(1)$ will be vary linearly with $Y_i$ (and be independent of small changes to $Y_s$), and thus $\gamma$ will be constant.
Within the $\eta^{0.5}$ and $\eta^{0.24}$ regimes for the tension and shear respectively, the displacement at the tip of the lamella is very much smaller than at its base, thus increasing $\eta$ serves to increase the ``penetration depth" of the deformation, and a complex interaction between the two materials creates the observed scaling laws. The value of $\eta$ at the transition from one regime to the other is thus dependent on aspect ratio of the structure. 
\vspace{5mm}

\subsection{Generation-2}
Here we establish the effect of adding secondary lamellae along the length of a generation-1 structure. 
We establish the mechanical response of the structure when the boundary is displaced under tension and under shear.
We also investigate the effect of changing the aspect ratio of both sets of lamellae, which are varied between 4 and 64 
The parameters used here are $l_{2,1} = 8\times 10^{-5}$m, $ t_{2,2} = 5\times 10 ^{-4}$m, $ t_i = 10^{-6}$m, $ t_s = 5\times 10^{-5}$m, $t_F = t_{2,1},$ $Y_i = 10^8$Pa, $\Delta = t_i/10$ and $\nu = 0.45$, other parameters are given by a specific aspect ratio.

\begin{figure}
\begin{center}
\resizebox{6.cm}{!}{\begingroup%
  \makeatletter%
  \providecommand\color[2][]{%
    \errmessage{(Inkscape) Color is used for the text in Inkscape, but the package 'color.sty' is not loaded}%
    \renewcommand\color[2][]{}%
  }%
  \providecommand\transparent[1]{%
    \errmessage{(Inkscape) Transparency is used (non-zero) for the text in Inkscape, but the package 'transparent.sty' is not loaded}%
    \renewcommand\transparent[1]{}%
  }%
  \providecommand\rotatebox[2]{#2}%
  \ifx\svgwidth\undefined%
    \setlength{\unitlength}{221.83027344bp}%
    \ifx\svgscale\undefined%
      \relax%
    \else%
      \setlength{\unitlength}{\unitlength * \real{\svgscale}}%
    \fi%
  \else%
    \setlength{\unitlength}{\svgwidth}%
  \fi%
  \global\let\svgwidth\undefined%
  \global\let\svgscale\undefined%
  \makeatother%
  \begin{picture}(1,0.71507939)%
    \put(0,0){\includegraphics[width=\unitlength,page=1]{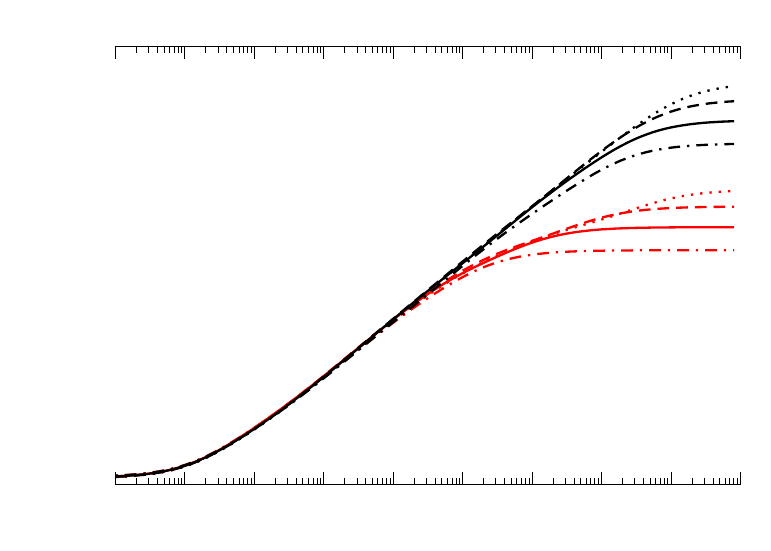}}%
    \put(0.14548516,0.04106583){\color[rgb]{0,0,0}\makebox(0,0)[lb]{\smash{1}}}%
    \put(0.2289963,0.04106583){\color[rgb]{0,0,0}\makebox(0,0)[lb]{\smash{10}}}%
    \put(0.30634979,0.04106583){\color[rgb]{0,0,0}\makebox(0,0)[lb]{\smash{100}}}%
    \put(0.39804753,0.04106583){\color[rgb]{0,0,0}\makebox(0,0)[lb]{\smash{$10^3$}}}%
    \put(0.48245144,0.04106583){\color[rgb]{0,0,0}\makebox(0,0)[lb]{\smash{$10^4$}}}%
    \put(0.57310448,0.04074116){\color[rgb]{0,0,0}\makebox(0,0)[lb]{\smash{$10^5$}}}%
    \put(0.663027,0.04074116){\color[rgb]{0,0,0}\makebox(0,0)[lb]{\smash{$10^6$}}}%
    \put(0.75335544,0.04106583){\color[rgb]{0,0,0}\makebox(0,0)[lb]{\smash{$10^7$}}}%
    \put(0.84352149,0.04106583){\color[rgb]{0,0,0}\makebox(0,0)[lb]{\smash{$10^8$}}}%
    \put(0.93352529,0.04106583){\color[rgb]{0,0,0}\makebox(0,0)[lb]{\smash{$10^9$}}}%
    \put(0,0){\includegraphics[width=\unitlength,page=2]{G2_amplification.pdf}}%
    \put(0.08962935,0.16117904){\color[rgb]{0,0,0}\makebox(0,0)[lb]{\smash{1}}}%
    \put(0.07623837,0.27723443){\color[rgb]{0,0,0}\makebox(0,0)[lb]{\smash{10}}}%
    \put(0.06511981,0.39320859){\color[rgb]{0,0,0}\makebox(0,0)[lb]{\smash{100}}}%
    \put(0.06088931,0.50910165){\color[rgb]{0,0,0}\makebox(0,0)[lb]{\smash{$10^3$}}}%
    \put(0.06387154,0.62507582){\color[rgb]{0,0,0}\makebox(0,0)[lb]{\smash{$10^4$}}}%
    \put(0,0){\includegraphics[width=\unitlength,page=3]{G2_amplification.pdf}}%
    \put(0.87540717,0.28813247){\color[rgb]{0,0,0}\makebox(0,0)[lb]{\smash{0.25}}}%
    \put(0.57889089,0.46376597){\color[rgb]{0,0,0}\makebox(0,0)[lb]{\smash{0.64}}}%
    \put(0,0){\includegraphics[width=\unitlength,page=4]{G2_amplification.pdf}}%
    \put(0.29814781,0.47242045){\color[rgb]{0,0,0}\makebox(0,0)[lb]{\smash{$a_{2,1} = 8$}}}%
    \put(0,0){\includegraphics[width=\unitlength,page=5]{G2_amplification.pdf}}%
    \put(0.29814781,0.42359988){\color[rgb]{0,0,0}\makebox(0,0)[lb]{\smash{$a_{2,1} = 16$}}}%
    \put(0,0){\includegraphics[width=\unitlength,page=6]{G2_amplification.pdf}}%
    \put(0.29814781,0.37477934){\color[rgb]{0,0,0}\makebox(0,0)[lb]{\smash{$a_{2,1} = 32$}}}%
    \put(0,0){\includegraphics[width=\unitlength,page=7]{G2_amplification.pdf}}%
    \put(0.29814781,0.3259588){\color[rgb]{0,0,0}\makebox(0,0)[lb]{\smash{$a_{2,1} = 64$}}}%
    \put(0.27484621,0.51611586){\color[rgb]{0,0,0}\rotatebox{31.55748461}{\makebox(0,0)[lb]{\smash{$a_{2,2} = 4$}}}}%
    \put(0.21256079,0.51989204){\color[rgb]{0,0,0}\rotatebox{30.25467644}{\makebox(0,0)[lb]{\smash{\textcolor{red}{$a_{2,2} = 64$}}}}}%
    \put(0.55643382,0.00206031){\color[rgb]{0,0,0}\makebox(0,0)[lb]{\smash{$\eta$}}}%
    \put(0.00213387,0.39594795){\color[rgb]{0,0,0}\makebox(0,0)[lb]{\smash{$\gamma$}}}%
    \put(0,0){\includegraphics[width=\unitlength,page=8]{G2_amplification.pdf}}%
    \put(-0.00104295,0.64671399){\color[rgb]{0,0,0}\makebox(0,0)[lb]{\smash{(a)}}}%
  \end{picture}%
\endgroup%
}
\resizebox{6cm}{!}{\begingroup%
  \makeatletter%
  \providecommand\color[2][]{%
    \errmessage{(Inkscape) Color is used for the text in Inkscape, but the package 'color.sty' is not loaded}%
    \renewcommand\color[2][]{}%
  }%
  \providecommand\transparent[1]{%
    \errmessage{(Inkscape) Transparency is used (non-zero) for the text in Inkscape, but the package 'transparent.sty' is not loaded}%
    \renewcommand\transparent[1]{}%
  }%
  \providecommand\rotatebox[2]{#2}%
  \ifx\svgwidth\undefined%
    \setlength{\unitlength}{220.40187988bp}%
    \ifx\svgscale\undefined%
      \relax%
    \else%
      \setlength{\unitlength}{\unitlength * \real{\svgscale}}%
    \fi%
  \else%
    \setlength{\unitlength}{\svgwidth}%
  \fi%
  \global\let\svgwidth\undefined%
  \global\let\svgscale\undefined%
  \makeatother%
  \begin{picture}(1,0.67689903)%
    \put(0,0){\includegraphics[width=\unitlength,page=1]{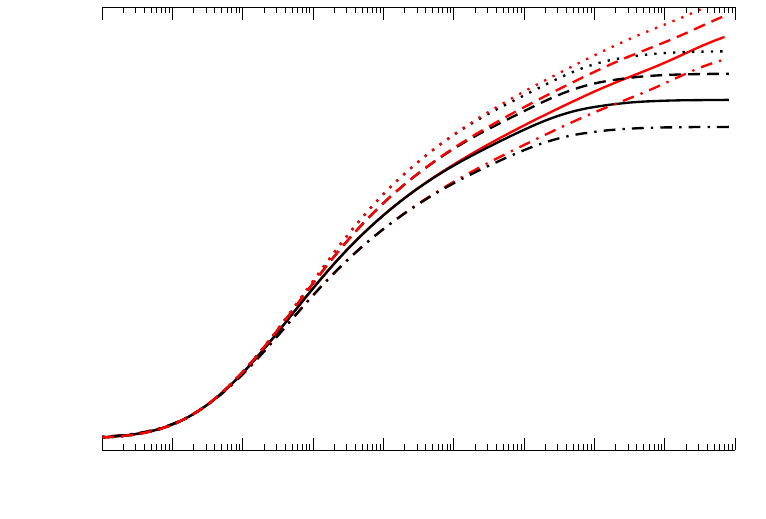}}%
    \put(0.1288908,0.04293829){\color[rgb]{0,0,0}\makebox(0,0)[lb]{\smash{1}}}%
    \put(0.21402372,0.04293829){\color[rgb]{0,0,0}\makebox(0,0)[lb]{\smash{10}}}%
    \put(0.30023215,0.04293829){\color[rgb]{0,0,0}\makebox(0,0)[lb]{\smash{100}}}%
    \put(0.38635787,0.04293829){\color[rgb]{0,0,0}\makebox(0,0)[lb]{\smash{$10^3$}}}%
    \put(0.47240086,0.04293829){\color[rgb]{0,0,0}\makebox(0,0)[lb]{\smash{$10^4$}}}%
    \put(0.56481432,0.04260736){\color[rgb]{0,0,0}\makebox(0,0)[lb]{\smash{$10^5$}}}%
    \put(0.65648317,0.04260736){\color[rgb]{0,0,0}\makebox(0,0)[lb]{\smash{$10^5$}}}%
    \put(0.7485657,0.04293829){\color[rgb]{0,0,0}\makebox(0,0)[lb]{\smash{$10^7$}}}%
    \put(0.84048279,0.04293829){\color[rgb]{0,0,0}\makebox(0,0)[lb]{\smash{$10^8$}}}%
    \put(0.93223438,0.04293829){\color[rgb]{0,0,0}\makebox(0,0)[lb]{\smash{$10^9$}}}%
    \put(0,0){\includegraphics[width=\unitlength,page=2]{G2_shear_amp.pdf}}%
    \put(0.08702922,0.14885681){\color[rgb]{0,0,0}\makebox(0,0)[lb]{\smash{1}}}%
    \put(0.07337818,0.3193708){\color[rgb]{0,0,0}\makebox(0,0)[lb]{\smash{10}}}%
    \put(0.0547842,0.48971936){\color[rgb]{0,0,0}\makebox(0,0)[lb]{\smash{100}}}%
    \put(0.0576377,0.66006791){\color[rgb]{0,0,0}\makebox(0,0)[lb]{\smash{$10^3$}}}%
    \put(0,0){\includegraphics[width=\unitlength,page=3]{G2_shear_amp.pdf}}%
    \put(0.69275494,0.26261115){\color[rgb]{0,0,0}\makebox(0,0)[lb]{\smash{$a_{2,1} = 64$}}}%
    \put(0,0){\includegraphics[width=\unitlength,page=4]{G2_shear_amp.pdf}}%
    \put(0.65528455,0.31333277){\color[rgb]{0,0,0}\rotatebox{38.10632331}{\makebox(0,0)[lb]{\smash{$a_{2,2} = 4$}}}}%
    \put(0.60280354,0.31699424){\color[rgb]{0,0,0}\rotatebox{38.10632331}{\makebox(0,0)[lb]{\smash{\textcolor{red}{$a_{2,2} = 16$}}}}}%
    \put(0,0){\includegraphics[width=\unitlength,page=5]{G2_shear_amp.pdf}}%
    \put(0.69275494,0.18954855){\color[rgb]{0,0,0}\makebox(0,0)[lb]{\smash{$a_{2,1} = 16$}}}%
    \put(0,0){\includegraphics[width=\unitlength,page=6]{G2_shear_amp.pdf}}%
    \put(0.69275494,0.15301718){\color[rgb]{0,0,0}\makebox(0,0)[lb]{\smash{$a_{2,1} = 8$}}}%
    \put(0,0){\includegraphics[width=\unitlength,page=7]{G2_shear_amp.pdf}}%
    \put(0.69275494,0.22607981){\color[rgb]{0,0,0}\makebox(0,0)[lb]{\smash{$a_{2,1} = 32$}}}%
    \put(0,0){\includegraphics[width=\unitlength,page=8]{G2_shear_amp.pdf}}%
    \put(0.30404636,0.31650723){\color[rgb]{0,0,0}\makebox(0,0)[lb]{\smash{0.7}}}%
    \put(0.52456786,0.54123886){\color[rgb]{0,0,0}\makebox(0,0)[lb]{\smash{0.25}}}%
    \put(-0.00143559,0.3814371){\color[rgb]{0,0,0}\makebox(0,0)[lb]{\smash{$\gamma$}}}%
    \put(0.54729809,0.00207361){\color[rgb]{0,0,0}\makebox(0,0)[lb]{\smash{$\eta$}}}%
    \put(-0.01595452,0.67181573){\color[rgb]{0,0,0}\makebox(0,0)[lb]{\smash{(b)}}}%
  \end{picture}%
\endgroup%
}
\caption{\footnotesize [Colour online] The stiffness increase of a generation-2 interface relative to a generation-0 geometry for various values of aspect ratio. The interface loaded under (a) tension and (b) shear. The results are shown for vairious structures where the aspect ratio of both the primary and secondary lamellae ($a_{2,2}$ and $a_{2,1}$ respectively) have been varied. }\label{G2_res}
\end{center}
\end{figure}

{\it Tension:} 
We first analyse the response of the structure when a displacement in the $-y$ direction is imposed on the base of the deformable surface, as indicated by the yellow dashed line in figure \ref{geometry}. As shown in figure \ref{G2_res}a, for the generation two structure three regimes are present:
For large $\eta$, the value of $\gamma$ is independent of $\eta$, this is indicative of a solid body translation of the material of stiffness $Y_s$.
For relatively small $\eta$, a novel scaling law is observed, 
\begin{equation}
\gamma \sim \eta^{0.64},\label{scaling_G2}
\end{equation}
where the error in the scaling is $\pm 0.02$. 
It is observed that this scaling law is dependent on the distance between primary lamellae on the deformable base and is maximised in the limit of small separation between primary lamellae. 
It can be shown (see supplementary information) that the maximum value of the power in Eq.~(\ref{scaling_G2}) is $0.66 \pm 0.01$.
For intermediate values of $\eta$ a secondary scaling is observed, 
\begin{equation}
\gamma \sim \eta^{0.25},
\end{equation}
where the error in the power is $\pm0.02$. 
In this regime, it is hypothesised that the primary lamella acts as a solid base and the structure becomes equivalent to a series of generation-1 lamellae with a shear deformation imposed at their bases (see previous section). 
It is noted that increasing the aspect ratio of the primary lamella (and therefore the value of $n_{2,1}$ for this geometry), serves to increase the minimum value of $\eta$ at which this scaling is observed, this supports the hypothesis that the whole of the primary lamellae becomes relatively stiff under tension resulting in this transition. 
\begin{figure}
\begin{center}
\resizebox{6.5cm}{!}{\begingroup%
  \makeatletter%
  \providecommand\color[2][]{%
    \errmessage{(Inkscape) Color is used for the text in Inkscape, but the package 'color.sty' is not loaded}%
    \renewcommand\color[2][]{}%
  }%
  \providecommand\transparent[1]{%
    \errmessage{(Inkscape) Transparency is used (non-zero) for the text in Inkscape, but the package 'transparent.sty' is not loaded}%
    \renewcommand\transparent[1]{}%
  }%
  \providecommand\rotatebox[2]{#2}%
  \ifx\svgwidth\undefined%
    \setlength{\unitlength}{297.07098293bp}%
    \ifx\svgscale\undefined%
      \relax%
    \else%
      \setlength{\unitlength}{\unitlength * \real{\svgscale}}%
    \fi%
  \else%
    \setlength{\unitlength}{\svgwidth}%
  \fi%
  \global\let\svgwidth\undefined%
  \global\let\svgscale\undefined%
  \makeatother%
  \begin{picture}(1,0.55932241)%
    \put(-0.00129855,0.29712333){\color[rgb]{0,0,0}\makebox(0,0)[lb]{\smash{$\frac{f_R(\theta)}{f_R(\pi/2)}$}}}%
    \put(0,0){\includegraphics[width=\unitlength,page=1]{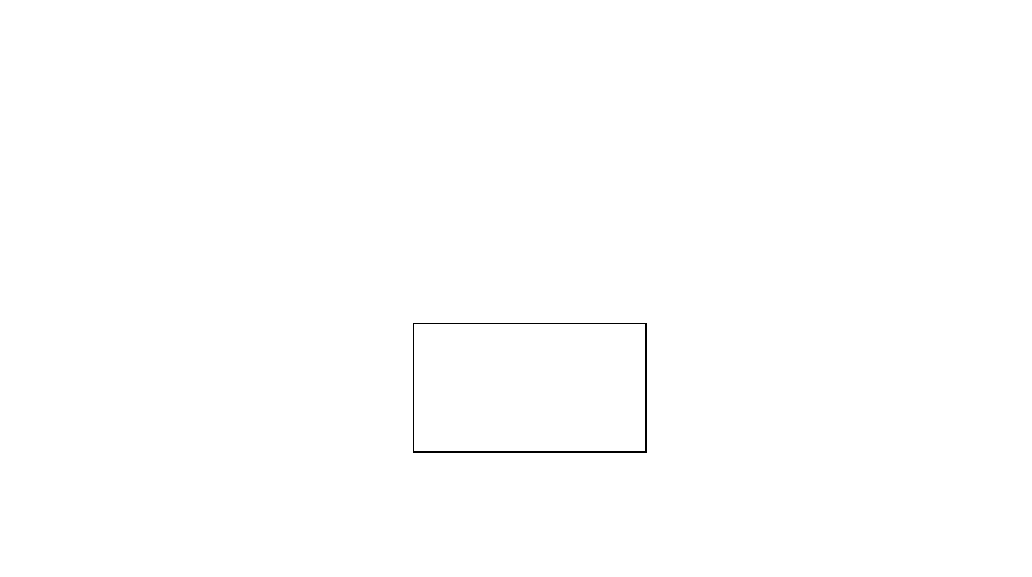}}%
    \put(0.475034628,0.13727993){\color[rgb]{0,0,0}\makebox(0,0)[lb]{\smash{$\eta = 10^{10}$}}}%
    \put(0,0){\includegraphics[width=\unitlength,page=2]{G2_tension_theta_only.pdf}}%
    \put(0.475034628,0.20632141){\color[rgb]{0,0,0}\makebox(0,0)[lb]{\smash{$\eta = 10^{4}$}}}%
    \put(0,0){\includegraphics[width=\unitlength,page=3]{G2_tension_theta_only.pdf}}%
    \put(0.475034628,0.17067178){\color[rgb]{0,0,0}\makebox(0,0)[lb]{\smash{$\eta = 10^{6}$}}}%
    \put(0,0){\includegraphics[width=\unitlength,page=4]{G2_tension_theta_only.pdf}}%
    \put(0.1119226,0.50352819){\color[rgb]{0,0,0}\makebox(0,0)[lb]{\smash{1.4}}}%
    \put(0.11209093,0.41666097){\color[rgb]{0,0,0}\makebox(0,0)[lb]{\smash{1.2}}}%
    \put(0.11194665,0.33225544){\color[rgb]{0,0,0}\makebox(0,0)[lb]{\smash{1.0}}}%
    \put(0.11248771,0.24746029){\color[rgb]{0,0,0}\makebox(0,0)[lb]{\smash{0.8}}}%
    \put(0.1119226,0.08466493){\color[rgb]{0,0,0}\makebox(0,0)[lb]{\smash{0.4}}}%
    \put(0.11201879,0.163675){\color[rgb]{0,0,0}\makebox(0,0)[lb]{\smash{0.6}}}%
    \put(0.60920392,0.04825417){\color[rgb]{0,0,0}\makebox(0,0)[lb]{\smash{$\frac{5\pi}{8}$}}}%
    \put(0.50523029,0.04825417){\color[rgb]{0,0,0}\makebox(0,0)[lb]{\smash{$\frac{\pi}{2}$}}}%
    \put(0.38373657,0.04825417){\color[rgb]{0,0,0}\makebox(0,0)[lb]{\smash{$\frac{3\pi}{8}$}}}%
    \put(0.27867278,0.04825417){\color[rgb]{0,0,0}\makebox(0,0)[lb]{\smash{$\frac{\pi}{4}$}}}%
    \put(0.7131777,0.04825417){\color[rgb]{0,0,0}\makebox(0,0)[lb]{\smash{$\frac{3\pi}{4}$}}}%
    \put(0.83358143,0.04825417){\color[rgb]{0,0,0}\makebox(0,0)[lb]{\smash{$\frac{7\pi}{8}$}}}%
    \put(0.49967129,0.00187567){\color[rgb]{0,0,0}\makebox(0,0)[lb]{\smash{$\theta$}}}%
    \put(0.16922221,0.04825417){\color[rgb]{0,0,0}\makebox(0,0)[lb]{\smash{$\frac{\pi}{8}$}}}%
    \put(0,0){\includegraphics[width=\unitlength,page=5]{G2_tension_theta_only.pdf}}%
  \end{picture}%
\endgroup%
}
\resizebox{6.5cm}{!}{\begingroup%
  \makeatletter%
  \providecommand\color[2][]{%
    \errmessage{(Inkscape) Color is used for the text in Inkscape, but the package 'color.sty' is not loaded}%
    \renewcommand\color[2][]{}%
  }%
  \providecommand\transparent[1]{%
    \errmessage{(Inkscape) Transparency is used (non-zero) for the text in Inkscape, but the package 'transparent.sty' is not loaded}%
    \renewcommand\transparent[1]{}%
  }%
  \providecommand\rotatebox[2]{#2}%
  \ifx\svgwidth\undefined%
    \setlength{\unitlength}{293.02477018bp}%
    \ifx\svgscale\undefined%
      \relax%
    \else%
      \setlength{\unitlength}{\unitlength * \real{\svgscale}}%
    \fi%
  \else%
    \setlength{\unitlength}{\svgwidth}%
  \fi%
  \global\let\svgwidth\undefined%
  \global\let\svgscale\undefined%
  \makeatother%
  \begin{picture}(1,0.55838309)%
    \put(-0.00129467,0.35318836){\color[rgb]{0,0,0}\makebox(0,0)[lb]{\smash{$\frac{f_{\mbox{R}}(\theta)}{f_{\mbox{R}}(\pi/2)}$}}}%
    \put(0.49910352,0.00187008){\color[rgb]{0,0,0}\makebox(0,0)[lb]{\smash{$\theta$}}}%
    \put(0,0){\includegraphics[width=\unitlength,page=1]{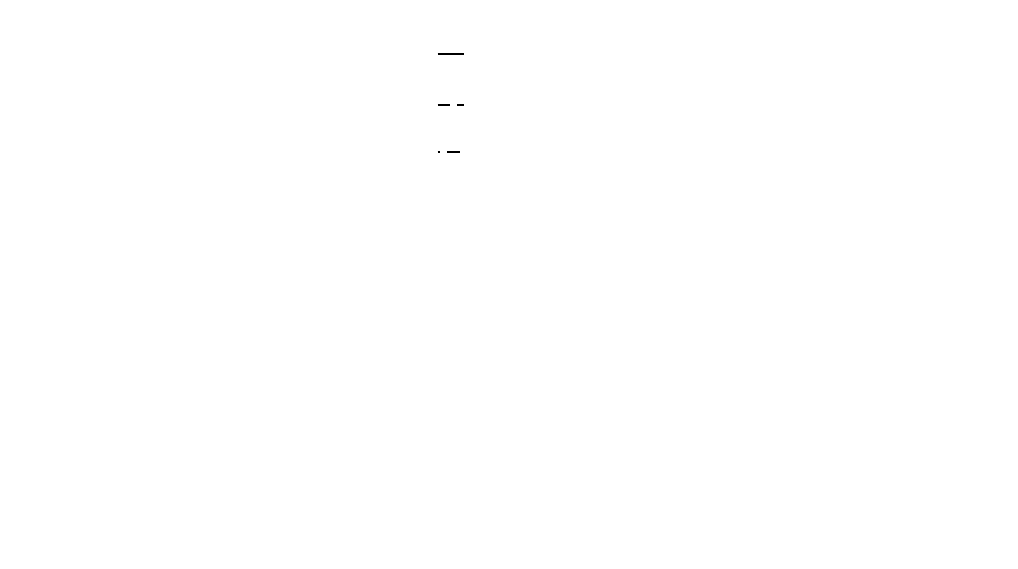}}%
    \put(0.12083696,0.0977056){\color[rgb]{0,0,0}\makebox(0,0)[lb]{\smash{0.75}}}%
    \put(0.15359864,0.19053043){\color[rgb]{0,0,0}\makebox(0,0)[lb]{\smash{1}}}%
    \put(0.12083696,0.27789513){\color[rgb]{0,0,0}\makebox(0,0)[lb]{\smash{1.25}}}%
    \put(0.12629727,0.36525983){\color[rgb]{0,0,0}\makebox(0,0)[lb]{\smash{1.5}}}%
    \put(0.11537665,0.46354492){\color[rgb]{0,0,0}\makebox(0,0)[lb]{\smash{1.75}}}%
    \put(0,0){\includegraphics[width=\unitlength,page=2]{G2_shear_theta_only.pdf}}%
    \put(0.47820809,0.39594763){\color[rgb]{0,0,0}\makebox(0,0)[lb]{\smash{$\eta = 10^{10}$}}}%
    \put(0.47820809,0.44498755){\color[rgb]{0,0,0}\makebox(0,0)[lb]{\smash{$\eta = 10^{6}$}}}%
    \put(0.47820809,0.49145134){\color[rgb]{0,0,0}\makebox(0,0)[lb]{\smash{$\eta = 10^{2}$}}}%
    \put(0.61583097,0.05645778){\color[rgb]{0,0,0}\makebox(0,0)[lb]{\smash{$\frac{5\pi}{8}$}}}%
    \put(0.51216741,0.05645778){\color[rgb]{0,0,0}\makebox(0,0)[lb]{\smash{$\frac{\pi}{2}$}}}%
    \put(0.39649665,0.05645778){\color[rgb]{0,0,0}\makebox(0,0)[lb]{\smash{$\frac{3\pi}{8}$}}}%
    \put(0.30266702,0.05645778){\color[rgb]{0,0,0}\makebox(0,0)[lb]{\smash{$\frac{\pi}{4}$}}}%
    \put(0.72495462,0.05645778){\color[rgb]{0,0,0}\makebox(0,0)[lb]{\smash{$\frac{3\pi}{4}$}}}%
    \put(0.83407816,0.05645778){\color[rgb]{0,0,0}\makebox(0,0)[lb]{\smash{$\frac{7\pi}{8}$}}}%
    \put(0.18262261,0.05645778){\color[rgb]{0,0,0}\makebox(0,0)[lb]{\smash{$\frac{\pi}{8}$}}}%
    \put(0,0){\includegraphics[width=\unitlength,page=3]{G2_shear_theta_only.pdf}}%
  \end{picture}%
\endgroup%
}
\caption{\footnotesize 
The variation in stiffness as a function of the angle at which the secondary lamellae protrude from the primary lamella, $\theta$ normalised by the stiffness of the geometry with $\theta = \frac{\pi}{2}$ for various values of $\eta$ with the interface loaded under tension (above) and shear (below). The three values of $\eta$ plotted are typical of the scaling regimes: $\gamma \sim \eta^{0.64}$ ($\gamma \sim \eta^{0.7}$ for shear), $\gamma \sim \eta^{0.25}$ ($\gamma\sim \eta^{0.24}$ for shear) and $\gamma$ independent of $\eta$.\label{G2_angle}}
\end{center}
\end{figure}
The variation of stiffness with the angle $\theta$ at which the secondary lamellae protrude from the primary lamella is shown in figure \ref{G2_angle}. Here a single lamella is separated from the base and the reaction force is measured when an imposed displacement along the green dashed line in figure \ref{geometry}. 
The parameters used here are: $a_{2,2} = 32, a_{2,1} = 32, t_i = 1\times10^{-4}\mbox{m}, t_{2,2} = 2\times 10^{-3}\mbox{m}, Y_i = 100\mbox{MPa}, \Delta = t_i/10, \nu = 0.45 \mbox{ and } t_F = 1.4\times 10^{-3}\mbox{m}$.
It is observed the the dependence on $\theta$ is dependent on the relative stiffness of the two deformable materials involved. 
In the regime where stiffness scales with $\eta^{0.64}$, it is seen that the minimum stiffness of the structure is obtained when $\theta = \frac{\pi}{2}$, indicating in this regime it is advantageous to orient the secondary lamellae such that the maximum proportion of the interface material is loaded under shear. 
In the other two scaling regimes, the reverse holds, $\theta = \frac{\pi}{2}$ is shown to correspond to maximum stiffness of the interface.

{\it Shear:}
The generation-2 interface can also be loaded under shear.
Here we impose a deformation of magnitude $\Delta$ in the $x$-direction. 
Figure \ref{G2_res}b shows a similar cascade of scaling laws is observed when the interface is loaded under tension. For $\eta \gg 1$, the increase in stiffness over a generation-0 interface is independent of $\eta$, this represents the regime where the whole generation-2 structure of stiffness $Y_s$ behaves as a rigid body. For small values of $\eta$, it is observed that 
\begin{equation}
\gamma \sim \eta^{0.7},
\end{equation}
where the error in the scaling is $\pm 0.02$. 
This scaling is dependent on the distance between lamellae and it is maximal observed value is $0.7$, see supplementary information. 
In this regime variations in $\eta$ serve to increase the penetration depth of the deformations in both the primary and secondary lamellae.
In the intermediate regime, the scaling of 
\begin{equation}
\gamma \sim \eta^{0.25},
\end{equation}
is observed (error in power is $\pm0.01$).
It is hypothesised that in this scaling regime the secondary lamellae behave as rigid beams under a compressive/tensile load. 
The transition from $\gamma$ scaling as $\eta^{0.7}$ and $\eta^{0.25}$ is dependent only on the aspect ratio of the secondary lamellae, supporting this hypothesis. 

The dependence of $f_R$ on $\theta$ is also established and shown in figure \ref{G2_angle}, we find for large values of $\eta \,(> 10^{8})$, increased stiffness of interface is obtained for values of $\theta$ close to 0 and $\pi$. For relatively small values of $\eta \;(\approx 10^{2}$) it is found that the stiffness is maximum for $\theta = \frac{\pi}{2}$, however, gain is less significant. These trends are the reverse to those observed for an interface under tension. It is notable however, that the results under shear are less symmetric about the point $\theta = \frac{\pi}{2}$ than those obtained under tension (figure \ref{G2_angle}).

\section{Discussion}

Here we have shown that the geometry observed in the equine hoof is conducive to a stiff interface between two materials of vastly differing stiffness. 
We have investigated two different loading conditions on the interface and found in both cases hierarchy can be utilised in a highly beneficial manner. 
We have shown that through manipulation of the number of degrees of hierarchy, the scaling laws relating stiffness of interface to adhesion stiffness can be manipulated in a non-trivial, systematic manner. 
Furthermore, we have shown through altering the aspect ratio of the lamellae, the value of $\eta$ at the transition from one scaling regime to another can be manipulated. 
Due to these altered scaling relationships, with increased hierarchy a greater degree of robustness under perturbation of the parameter $\eta$ relating the two material stiffness is obtained, thus offering a possible evolutionary advantage of the hierarchical design over other designs that create a similar interface stiffness for a given $\eta$ value. 

This work shows the great potential of geometry to dictate the mechanical behaviour of interfaces joining two dissimilar materials. Although higher degrees of hierarchy prove too computationally expensive to simulate directly, analytic studies should be undertaken to elucidate the behaviour of these geometries. Though this work shows the tailorability of interface stiffness through hierarchy open questions remain regarding the fracture resistance, ductility and strength of these intricate architectures. 

Not only does this work have implications for the design of geometry in engineering porblems of joining two materials of differing stiffness, this work also lends credence to the long assumed role of hierarchical interface geometry in the equine hoof: that such geometry has been selected by evolutionary pressures (among other reasons) for mechanical purpose. 
Despite the undoubted complexity of the biological system, it is expected that the main computational findings of this work apply to the interface observed within the hoof: increasing the hierarchy of the interface serves to increase the mechanical stiffness of the interface. 
Further work is necessary to establish the effect of this interface on fracture resistance and maximum allowable loads and the effect of higher degrees of hierarchy. 

\section{Acknowledgments}
The authors would like to thank Catrin Rutland and Ramzi Al-Agele. DRK would like to acknowledge funding support from the Academy of Finland.


\section{Bibliography}

\begin{thebibliography}{1}

\bibitem{Federle_2006} Federle W., J. Exp. Biol. {\bf 209} 2611 (2006)
\bibitem {Gao_2006} Yao H. and Gao H., Journal of the Mechanics and Physics of Solids {\bf 54} 1120 (2006).
\bibitem{Labonte_2016} Labontea D., et al., Proc. Natl Acad. Sci. USA {\bf 113}, 1297 (2016)
\bibitem{Huber_2005} Huber, G., et al., Proc. Natl Acad. Sci. USA {\bf 102}, 293 (2005)
\bibitem{Autumn_2002} Autumn, K., et al., Proc. Natl Acad. Sci. USA {\bf 99} 12252 (2002)
\bibitem{Cutkosky_2015} Cutkosky, M. R., et al., Interface Focus, {\bf 5} 20150015 (2015)
\bibitem{Geim_2003} Geim, A. K., et al., Nature Materials {\bf 2}, 461 (2003)
\bibitem{Brodoceanu_2016} Brodoceanu, D., et al., Bioinspir. Biomim. {\bf 11}, 051001 (2016)
\bibitem{Autumn_2008} Autumn, K. and Gravish, N., Phil. Trans. R. Soc. A {\bf 366} 1575 (2008)
\bibitem{Zhou_2005} Zhou, H. and Zhang, Y., Phys. Rev. Lett. {\bf 94}, 028104 (2005)
\bibitem{Ritchie_2011} Ritchie, R. O., Nat. Mater. 10, 817–822 (2011)
\bibitem{Li_2012} Li, Y, et al., Phys. Rev. E, {\bf 85}, 031901 (2012)
\bibitem{Huiskes_2000} Huiskes, R.,et al., Nature (London) {\bf 405}, 704 (2000).
\bibitem{Schaelder_2011} T. A. Schaelder, et al., Science 334, 962 (2011)
\bibitem{Lin_2014} E. Lin et al., J. Mat. Res. {\bf 9}, 1867 (2014)
\bibitem{Rayneau-Kirkhope_2013} Rayneau-Kirkhope, D., et al., Phys. Rev. E, {\bf 187}, 26244 (2013)
\bibitem{Rayneau-Kirkhope_2012} Rayneau-Kirkhope, D., et al., Phys. Rev. Lett. {\bf 109} 204301 (2012)
\bibitem{Li_2014} Lin, E., et al., J. Mech. Phys. Solids {\bf 73}, 166 (2014)
\bibitem{Li_2012_a} Li, Y., et al., Phys. Rev. E, {\bf 84}, 062904 (2012)
\bibitem{Boyce_2013} Y. Li, et al., J. Mech. Phys. Solids {\bf 61}, 1144-1167 (2013)
\bibitem{Shahar_2009} Krauss, S., et al., Adv. Mat. {\bf 21} 407 (2009)
\bibitem{Gao_2006b} Gao, H. J., Int. J. Fracture {\bf 138} 101 (2006)
\bibitem{Hubbard_1971} Hubbard, R. P. et al., J. Biomechanics {\bf 4}, 491 (1971)
\bibitem{Kondo_2016} Inoue, S. and Kondo, S. Scientific Reports, {\bf 6}, 33689 (2016)
\bibitem{Boyce_2012} Li, Y. et al., Phys. Rev. E, {\bf 85}, 031901 (2012)
\bibitem{Stump_1967} Stump, J. E., J Am Vet Med Assoc {\bf 151} 1588 (1967)
\bibitem{Kainer_1989} Kainer, R. A., Vet Clin North Am Equine Pract {\bf 5} 1 (1989)
\bibitem{Pollitt_2004} Pollitt, C. C., Clinical Techniques in Equine Practice {\bf 3} 3 (2004)
\bibitem{Pollitt_1996a} Pollitt, C. C., Equine Vet Educ {\bf 10} 318 (1996)
\bibitem{Pollitt_1999} Pollitt, C. C., AAEP Proc. {\bf 45} 188 (1999)
\bibitem{Pollitt_2010} Pollitt, C. C., The Veterinary clinics of North America Equine practice {\bf 26} 29 (2010).
\bibitem{COMSOL} COMSOL Inc., http://www.comsol.com/comsol-multiphysics.
\end{thebibliography}
\end{document}